\documentclass[12pt]{article}
\begin{document}

\addtolength{\baselineskip}{0.5\baselineskip}

\title{\textbf{Exploring the Harmony between Theory and Computation \\
- Toward a unified electronic structure theory}}
\author{Liqiang Wei \\
Institute for Theoretical Atomic, Molecular and Optical Physics\\
Harvard University, Cambridge, MA 02318}

\maketitle

\begin{abstract}
\vspace{0.05in}

The physical aspect of a general perturbation theory is explored.
Its role as a physical principle for understanding the interaction
among the matters with different levels of hierarchy is
appreciated. It is shown that the general perturbation theory can
not only be used for understanding the various electronic
phenomena including the nature of chemical bonds but also serve as
a unified theme for constructing general electronic structure
theories and calculation schemes.
\end{abstract}

\vspace{0.35in}

Perturbation theory is regarded as one of the two major approaches
for approximately solving quantum many-body problems. However, its
deeper physical aspect is far more than it is currently being used
just as a mathematical tool for solving the complicated issues.
All the fundamental laws in physics are variational in nature,
including the Schr$\ddot{o}$dinger equation in quantum mechanics.
Nevertheless, the perturbation theory provides a principle that
governs how the matters with different levels of hierarchy
interact. In fact, a general perturbation theory itself contains
two ingredients. On one hand, the degenerate or near-degenerate
situation is $\it{not}$ a perturbation at all but actually
constitutes a strong physical interaction. On the other hand, the
non-degenerate case is a real perturbation in the common sense. We
believe that, it is this equal or near energy physical mixing that
governs the interaction among the matters with different levels of
hierarchy. Of course, it is also the physical principle based on
which a unified chemical bond theory can be built.

Electrons are quantum mechanical particles which possess
wave-particle duality. The binding process of the electrons
associated with some atoms, or equivalently, the interaction of
atomic orbitals for the formation of a molecule, can be regarded
as a wave interference phenomenon. The interaction of
$\it{intra}$-atomic orbitals with the same energy or near energies
is the Pauling's hybridization process, which determines the
direction of chemical bonds, while the interaction of
$\it{inter}$-atomic orbitals with the same energy or near energies
determines the actual formation of chemical bonds. These are the
nature of chemical bonds [1,2]. The immediate benefit for
recognizing this near energy principle in determining the
formation of chemical bonds is that it gives a better
understanding of many previously developed very important
structure concepts such as multi-center chemical bonds, multiple
chemical bonds, resonance structure, Walsh diagrams, and avoided
crossing, and therefore incorporate them into one
$\it{qualiatative}$ theoretical framework [1,3-5].

More important in realizing this fundamental physics for
understanding how the matters interact is that it also provides a
physical foundation for $\it{quantitatively}$ investigating the
electronic structure of molecules, including large systems such as
molecular materials and biomolecules. We are going to have a
harmony between theory and computation.

$\underline{\it{Energy\ scale\ principle\ in\ Rayleigh-Ritz\
variatioanl\ approach}}$

Rayleigh-Ritz variational method is most commonly used for solving
eigenvalue problem in quantum mechanics. Its relation to the
general perturbation theory, including the degenerate situation,
has also been worked out mathematically during the 1960's [6].
However, the physical implication of this relation, especially its
role as a guidance in constructing the electronic structure
calculation schemes has not been explored and appreciated yet.
First of all, as long as the reference Hamiltonian which produces
the basis functions is made as close as possible to the full
Hamiltonian, then the dimension for the Rayleigh-Ritz variational
expansion will be made as small as possible. Secondly, if the
basis functions have the closest energies of the reference
Hamiltonian, then they will have the strongest mixing and make the
greatest contribution to the combined state, while the others with
larger energy differences will have smaller or even negligible
contributions. These are the situations we qualitatively discussed
above for the general perturbation theory. We term this as the
energy scale principle in Rayleigh-Ritz variational approach.

(a) $\underline{\it{Molecular\ fragmentation\ and\ basis\ set\
construction}}$

The basis set approach is a most popular and natural way for
solving the single particle equation such as Hartree-Fock
equation. Physically, it reflects a composite relation between the
molecule and its constituent atoms. To have an overall accurate
electronic structure calculation, the first necessary step is to
get the reliable and converged molecular orbitals [7].

However, since the current basis functions like most commonly used
contracted Gaussians are primarily a reflection of electrons in
single atoms in the molecule, it leaves the perturbed part of the
molecular Fock operator very large. That is why the polarization
functions, including some expanded ones such as the correlation
consistent basis sets, have to be introduced to get good
computation results [8]. Nevertheless, the $O(N^{4})$ scaling,
where $N$ is the number of basis functions, has become a major
bottleneck in quantum chemistry calculation, especially for the
large systems.

To overcome this difficulty, the energy-scale principle described
above can come for a help. If we construct the basis functions
which are the reflection of molecular fragments so that the
corresponding reference Hamiltonian is as close as possible to the
whole molecular Fock operator, then the dimension of basis set
expansion can be made as small as possible. This is going to be a
challenge work but will be mathematical in nature. The basis set
superposition effects (BSSE) is an example [9].

Similar situation occurs in the quantum molecular scattering
calculation, where the channels are used as the basis functions
for solving the Schr$\ddot{o}$dinger equation, or its integral
form, Lippmann-Schwinger equation with proper boundary conditions.
Since there are often very large differences between the channels
and the scattering waves for the whole reactive system in the
interaction regions, the dimension for their expansion is
particularly large, which causes the quantum scattering
calculation prohibitively expansive for all but the smallest
systems. The ideas suggested here can obviously be utilized for
remedying this deficiency.

(b)${\underline{\it{General\ multi-reference\ electronic\
structure\ theory}}}$

To get final accurate solution to the Schr$\ddot{o}$dinger
equation for the many-electron systems with a non-separable
two-body Coulomb interaction, it is most likely that we have to go
beyond the single particle description [10-13]. Mathematically,
the full configuration interaction (FCI) gives exact answers [14].
However, it is computationally prohibitive and possibly will never
been strictly realized. The energy scale principle described above
can also be applied in this $\it{configuration}$ level. A general
electronic structure theory should be multi-configuration or
multi-reference in nature [15-20]. First, there exists a strong
configuration mixing, for example, at transition states, for
excited states, and for multiple chemical bonds. The concept of
exciton introduced in solid state physics also belongs to this
case [21]. Second, the degenerate configurations are often the
case for the stable open-shell systems. Third, if we want to treat
the ground state and the excited states simultaneously, we have to
include the corresponding reference states in the same model
space. Finally, the separation of correlation into the static and
dynamic parts, which corresponds to the near degenerate and the
perturbed situations, really has chemical structure signature.
Therefore, among all the correlation approaches developed so far
for electronic structure, the MCSCF type with perturbation or
coupled-cluster expansion correction should be the most
appropriate and general one and works in the right direction. To
solve the remaining issues such as proper selection of
configurations for the model space, the efficient treatment of
dynamic correlation, and the avoidance of intruder states, we not
only need a mastery of current quantum many-body theory but also
might need its further development.

The importance and necessity of separation of correlation into a
static part and a dynamic part is also indicated in the DFT
calculation for the highly charged ions [22] and in its treatment
of transition states of reactions [23]. This calls for an
extension of current DFT to incorporate the differentiation of
static and dynamic correlation effects into its theoretical
framework [24,25].

(c)${\underline{\it{General\ pseudopotential\ theory}}}$

The concepts of pseudopotentials, effective core potentials (ECP),
or model potentials (MP) are those of the most significant
developments in the fields of electronic structure for molecular
and solid state systems. It treats valence electrons only, leaving
the core electrons and nucleus as a whole charge entity and
therefore reducing the number of electrons as well as the
corresponding overall size of the basis set being used for the
computation. It is important when we study the electronic
structure for large molecules or inorganic molecules containing
heavy elements [26]. A most commonly used pseudopotential for
solid state calculation is the so-called norm conserving
pseudopotential [27]. In addition to having the same valence state
energies, its pseudo valence state wavefunctions are also
equivalent to the valence state wavefunctions obtained from the
full electron calculations outside a cutoff radius. The
pseudopotentials constructed in this manner share the same
scattering properties as those of the full potentials over the
energy range of the valence states. The practical implementation
of various pseudopotentials has also demonstrated the importance
of choosing a correct size of the core or range of the valence
electrons for the accurate pseudopotential computation in order
that the core-valence correlations or core polarization can be
neglected. Obviously, the physics behind this valence and core
state separation is the energy scale principle we described above
applied in the level of $\it{atomic\ orbitals}$. After realizing
this principle, however, we might establish a more general
pseudopotential theory. We are planning to reformulate the
pseudopotential approach in the framework of perturbation theory
so that most flexible and accurate ECPs can be developed. They can
be used in different chemical environments and work for both
ground and excited state problems. The final goal is to make the
effective core potentials to be a routine rather than an
approximation for calculating electronic structure for large
molecules, inorganic molecules containing heavy elements, and
solid state systems.

(d)${\underline{\it{Molecular\ fragmentation\ and\ combined\
QM/MM\ approach\ for}}}$ \newline ${\underline{\it{electronic\
structure\ of\ large\ molecules}}}$

Combined QM/MM approach has become very popular in recent years in
the study of, for example, the chemical reactions in solutions and
in enzymes [28,29]. The basic consideration is that treating a
full collection of electrons for the whole system explicitly is
not only unrealistic but also unnecessary. In the first place, the
electronic charge redistribution induced by a chemical reaction is
very often limited to a small region due to the length scale
issues such as finite range of interaction or natural charge
distribution. Second, the quantum exchange effect for the
electrons is finite range, and there is no exchange interaction
among the electrons with long distance. This permits a partition
of the whole system into an active part and an inactive part
without any charge redistribution. The former has to be described
quantum mechanically since it possibly involves bond breaking and
making, while the latter can be described by molecular mechanics
because it merely serves as a classical electrostatic environment
for the active site. This combined QM/MM description has shown
remarkable successes in studying the electronic structure and
reactivity of large molecules in recent years. However, challenges
remain. One of the major obstacles for the applications is in the
proper treatment of boundary region where the cut has to be for a
covalent bond. Currently, there are two approaches to this
problem. The one introducing link atoms along the boundary is
severely limited and cannot be applied to treat a large variety of
different chemical systems. In addition, it artificially brings
additional forces into the system and therefore complicates the
problem. The other kind like local self-consistent field methods
seems reasonable but it is still more empirical. In order to
utilize this kind of combined QM/MM methods for investigating the
electronic structure and molecular dynamics in a larger domain of
fields, we need to develop a more generic ab initio approach. We
believe that the energy scale principle discussed above can play a
key role here. It is not only the principle according to which the
atomic orbitals including valence ones interact along the boundary
but also the principle based on which a systematic approach for
constructing the correct charge distribution or the force fields
along the boundary can be established. This is also the key for a
more sophisticated or finer treatment of quantum region including
its electron correlation.

In summary, the energy scale principle for the hierarchy of
interacting matters is identified. It not only can be utilized as
a general principle for understanding how the matters interact at
different levels but also can serve as the foundation based on
which the accurate electronic structure calculation schemes for
even large molecular systems can be constructed. It can also be
employed to build $\underline{\it{a\ general\ theory\ for\ the\
intermolecular\ forces}}$ so that the important issues such as the
interplay between chemical bondings and intermolecular forces can
be investigated [30].

\end{document}